\documentclass{mscs}
\usepackage{latexsym,amssymb,amsfonts,amsmath}

\title[Physics and computer science: quantum computation and other approaches]{Physics and computer science: quantum computation and other approaches}
\author[S.E. Venegas-Andraca]{S\ls A\ls L\ls V\ls A\ls D\ls O\ls R\ns E.\ns V\ls E\ls N\ls E\ls \ls G\ls A\ls S-A\ls N\ls D\ls R\ls A\ls C\ls A$^1$\\%
$^1$ Quantum Information Processing Group, 
\addressbreak 
Mathematics Department, 
Faculty of Engineering, 
\addressbreak 
Tecnol\'{o}gico de Monterrey Campus Estado de M\'{e}xico. 
\addressbreak 
Carretera Lago Gpe. Km 3.5, Atizap\'{a}n de Zaragoza, Edo. M\'{e}xico, M\'{e}xico. 
\addressbreak 
e-mail: 	salvador.venegas-andraca@keble.oxon.org}

\date{.}
\begin{document}
\maketitle

Computer science and computer engineering are disciplines that have definitely permeated and transformed every aspect of modern society. In these fields,  cutting-edge research is about new models of computation, new materials and techniques for building computer hardware, novel methods for speeding-up algorithms, and building bridges between computer science and several other scientific fields, bridges which allow scientists to both think of natural phenomena as computational procedures as well as to employ novel models of computation to simulate natural processes (for example, quantum walks have been used to model energy transport in photosynthetic light harvesting complexes \cite{hoyer10,caruso10}). A convergence of scientific, technological, economic, and epistemological demands is driving and integrating this research.  


Now, the theory of computation, in its canonical form, has a severe drawback: it does not take into account the physical properties of those devices used for performing computational or information processing tasks. Recent research approaches have therefore concentrated on thinking of computation
in a physical context. The rationale behind this approach is as follows: it must be possible to build the notions of information and computation upon physical principles because the behaviour of any device used for computation or  information processing must be, ultimately, predicted by the laws of physics. 
Among those physical theories that could be used for this purpose, quantum mechanics stands in first place.

Quantum mechanics and the theory of computation are two of the most important intellectual achievements of the 20th century. These two branches of science have not only inspired several generations of scientists and thinkers, they have also had a significant impact in the daily life of Mankind, from war to literature. As a matter of fact, cross-fertilisation between physics and computation has been abundant due to the fact that many ideas, concepts and technological developments from both fields have been used to advance knowledge in each discipline.

\newpage{}
Quantum Computation, one of the most recent joint ventures between physics and computer science, can be defined as the scientific field whose purpose is to solve problems with algorithms that exploit the quantum mechanical properties of those physical systems used to implement them. Among the theoretical discoveries and promising conjectures that have positioned quantum computation as a key element in modern science, we find:  1) the development of novel and powerful methods of computation 
that may allow us to significantly increase our processing power for solving certain problems \cite{nielsen00,gruzka00,kitaev02,lanzagorta09,childs10,jordan05} and 2) the simulation of complex physical systems \cite{feynman82, harris10,brown10}. A detailed summary of scientific and technological applications of quantum computers can be found in \cite{roadmapusa04,qist07}.

As for the physical realisation of quantum computers, several experimental platforms have been developed or customized over the last two decades. Indeed, although it is too early to predict the winning technologies for the implementation  of quantum computers, encouraging advances have been made over the last few years in fields  like quantum optics \cite{joo06,lanyon07,prevedel07,branderhorst08}, ion traps \cite{porras06,benhelm08}, and diamond-based technology \cite{weber10}. Moreover, according to the quantum computation roadmaps produced in the United States of America in 2004 \cite{roadmapusa04} and the European Union in 2007 \cite{qist07}, it is reasonable to expect quantum hardware with enough number of qubits and fault tolerant error correction ready to run quantum simulation and some quantum algorithms by 2012-2017.  

In addition to the development of quantum algorithms, there are other approaches towards building algorithms inspired in (classical) physical theories. Both quantum and classical approaches are inmmensely valuable to computer scientists as they provide us with new insights with respect to what a computer can ultimately do and how we can achieve it.

Thus, in the history of cross-fertilisation between physics and computation, this special volume on quantum algorithms and other physics-inspired algorithms is meant to be situated as a contribution towards realising what Nature has to say with respect  to our capacity to build automatic procedures for problem solving. 

I would like to finish this introduction by warmly thanking Professor Giuseppe Longo, chief editor of Mathematical Structures in Computer Science, for giving me the opportunity to edit this special volume as well as for his never-ending patience and support. Also, I am most grateful with the authors of this special volume papers for their outstanding scientific contributions and their hard work. It am deeply honoured for having been given the opportunity to work with you all. 


\newpage{}


\begin{thebibliography}{}

\bibitem[Hoyer et al 2010]{hoyer10} 
Hoyer, S., Sarovar, M., Whaley, K.B. (2010) 
Limits of quantum speedup in photosynthetic light harvesting.
\emph{New J. Phys.} \textbf{12}, 065041. 

\bibitem[Caruso et al 2010]{caruso10} 
Caruso, F., Chin, A.W., Datta, A., Huelga, S.F., Plenio, M.B.  (2010) 
Entanglement and entangling power of the dynamics in light-harvesting complexes.
\emph{Phys. Rev. A} \textbf{81}, 062346. 

\bibitem[Nielsen and Chuang 00]{nielsen00}
Nielsen, M.A., Chuang, I.L. (2000)
Quantum Computation and Quantum Information.
\emph{Cambridge University Press}.

\bibitem[Kitaev et al 2002]{kitaev02}
Kitaev, A.Y., Shen, A.H., Vyalyi, M.N. (2002)
Classical and Quantum Computation.
\emph{American Mathematical Society}.

\bibitem[Gruzka 2000]{gruzka00}
Gruzka, J. (2000)
Quantum Computation.
\emph{McGraw Hill}.

\bibitem[Lanzagorta and Ullman 2009]{lanzagorta09}
Lanzagorta, M., Ullman, J. (2009)
Quantum Computer Science.
\emph{Morgan and Claypool Publishers}.

\bibitem[Childs and van Dam 2010]{childs10}
Childs, A., van Dam, W. (2010)
Quantum algorithms for algebraic problems.
\emph{Rev. Mod. Phys.} \textbf{82}, 1-51. 

Stephen P. Jordan
\bibitem[Jordan 2005]{jordan05}
Jordan, S. (2005)
ast quantum algorithm for numerical gradient estimation.
\emph{Phys. Rev. Lett} \textbf{95}, 050501. 


\bibitem[Feynman 1982]{feynman82} 
Feynman, R.P. (1982)
Simulating Physics with Computers.
\emph{International Journal of Theoretical Physics} \textbf{21(6/7)}, 467-488. 

\bibitem[Harris and Kendon 2010]{harris10} 
Harris, S., Kendon, V.M. (2010) 
Quantum-assisted biomolecular modelling.
\emph{Phil. Trans. R. Soc. A} \textbf{368(1924)}, 3581-3592. 

\bibitem[Brown et al 2010]{brown10} 
Brown, K.L., Munro, W.J., Kendon, V.M. (2010) 
Using Quantum Computers for Quantum Simulation.
\emph{arXiv:1004.5528}. 

\bibitem[QIST 2007]{qist07}
ERA-Pilot (2007)
Quantum Information Processing and Communication Strategic Report version 1.5.
\emph{http://www.qist-europe.net/}

\bibitem[QIST 2004]{roadmapusa04}
QIST 2004. Advanced Research and Development Activity (2004)
A Quantum Information Science and Technology Roadmap, USA.
\emph{http://qist.lanl.gov/}

\bibitem[Joo et al 2006]{joo06}
Joo, J., Lim, Y.L., Beige, A., Knight, P.L. (2006)
Single-qubit rotations in 2D optical lattices with multi-qubit addressing.
\emph{Phys. Rev. A} \textbf{74}, 042344. 

\bibitem[Lanyon et al 2007]{lanyon07}
Lanyon, B.P., Weinhold, T.J., Langford, N.K., Barbieri, M., James, D.F.V., Gilchrist,  A., White, A.G. (2007)
Experimental Demonstration of a Compiled Version of Shors Algorithm with Quantum Entanglement.
\emph{Phys. Rev. Lett.} \textbf{99}, 250505. 

\bibitem[Prevedel et al 2007]{prevedel07}
Prevedel, R., Walther, P., Tiefenbacher, F., B{\"{o}}hi, P., Kaltenbaek, R., Jennewein, T., Zeilinger, A. (2007)
High-speed linear optics quantum computing using active feed-forward.
\emph{Nature} \textbf{445}, 65-69. 

\bibitem[Benhelm et al 2008]{benhelm08}
Benhelm, J., Kirchmair, G., Roos, C. F., Blatt, R. (2008)
Towards fault-tolerant quantum computing with trapped ions.
\emph{Nature Physics} \textbf{4}, 463. 

\bibitem[Branderhost et al 2008]{branderhorst08}
Branderhorst, M.P.A., Londero, P., Wasylczyk, P., Brif, C., Kosut, R.L., Rabitz, H., Walmsley, I.A. (2008)
Coherent control of decoherence.
\emph{Science} \textbf{320(5876)}, 638-643. 

\bibitem[Porras et al 2006]{porras06}
Porras, D., Cirac, J.I. (2006)
Quantum Manipulation of Trapped Ions in Two Dimensional Coulomb Crystals.
\emph{Phys. Rev. Lett.} \textbf{96}, 250501. 

\bibitem[Weber 2010]{weber10} 
Weber, J. R., Koehl, W. F.,  Varley, J. B., Janotti, A., Buckley, B. B., Van de Walle, C. G., Awschalom, D. D. (2010) 
Quantum computing with defects.
\emph{PNAS} \textbf{107 (19)}, 8513-8518. 




\end{thebibliography}
\end{document}